# Processing of Communication Signal Using Operational Transconductance Amplifier

A. Roy, K. Ghosh, S. Mondal and B. N. Ray

**Abstract**— This paper proposes a signal processing methodology of communication system and realized that circuits using operational transconductance amplifier (OTA). Two important classes of communication circuit, delta modulator and compander have been designed using that procedure. In the first implementation coded pulse modulation system is demonstrated which employ sampling, quantizing and coding to convert analog waveforms to digital signals while the second gives data compression and expansion in digital communication system. The proposed compander circuit is realized with operational transconductance amplifier and diode. Required power supply to operate the circuit is 3.5V. Performance of the circuits realized with OTAs has been demonstrated through SPICE simulation.

——————————  ◆  ——————————

## 1 INTRODUCTION

THE objective of this paper is to propose a signal processing methodology for communication system with a network of operational transconductance amplifier (OTA). Design procedure of two important class of communication system delta modulation and compander have been demonstrated. A novel technique to convert analog signal to digital signal is Delta Modulation. The difficulty of Linear Delta Modulator is removed in Adaptive Delta Modulator. In this paper both categories of Delta Modulator, realized with OTA have been studied. Data Compression is an essential part in communication system particularly in voice transmission [1]. Data compressor is a nonlinear device that compresses the signal amplitude. The range of voltage covered by voice signals, from loud talk to weak talk, is in the order of 1000 to 1. A non-uniform quantizer, characterized by a step size that increases as the separation from the origin of the transfer characteristic is increased, satisfies this requirement [1]. Non-uniform quantization can be practically implemented with a compressor followed by a uniform quantizer. The original signals are restored by the reverse process of compressor, known as expander. The combination of a compressor and an expander is called compander. A new design technique of low power OTA based compander circuit is introduced in this work. The motivation to undertake this research stems from the following considerations. (i) Growing popularity of analog and mixed signal ICs has provided the impetus to explore innovative designs implemented with CMOS VLSI technology and (ii) Reduction of design turn around time. Discrete passive components are major barrier to reduce area overhead in analog VLSI implementation.

OTA is an excellent current mode device to realize high frequency resistor-less analog designs. With this backdrop, we have embarked on design of OTA based RF communication circuits. A survey of the literature dealing with OTA based designs depicts following picture. A number of researchers have employed OTA as the basic building block in the design of non-linear networks. Edgar Sanchez-Sinencio et al. [2] have reported synthesis of a number of nonlinear circuits with OTA network. The authors have reported two synthesis techniques viz. rational approximation functions and piece-wise linear approximation technique. Four-quadrant multiplier and a phase shifter are the two important analog functions used in data communication system. Bang Sup - Song [3] has reported the recent development in respect of synthesis of these functions with OTA networks. Modulator circuit for analog and digital communication system (AM, ASK, FM, and FSK) has been synthesized and realized with single output OTA by Ray et al. [4]. Similar type of digital communication circuit (i.e., ASK/FSK/PSK/QAM) using multiple output OTA and a number of digitally controlled switches is proposed by Taher et al. [5].Also, Hietala proposes PSK/GMSK polar transreceiver [6], Rahim et al. present software defined wireless receiver [7]. It is well accepted that the design of the radio frequency (RF) section in a communication integrated circuit (IC) is a challenging problem.

In the recent past a few researchers have presented design methodology of compander circuit. In [8] Zrilic et.al proposes a compander circuit based on sigma delta modulator. That proposed circuit is implemented with fully digital devices. A new class of compander system is proposed that combines conventional broad band companders with adaptive filter is proposed in [9]. Sakran et al [10] present a new scheme based on μ law compander, where opti-

————————————————

- *A. Roy is with the Electronics and Telecommunication Department, Bengal Engineering and Science University, Shibpur, Howrah-711103, India.*
- *K. Ghosh is with the Department of Electronics & Communication Engineering, IERCEM Institute of Information Technology, India.*
- *S. Mondal is with the CST Department Bengal Engineering and Science University, India.*
- *B. N. Ray is with the Electronics and Telecommunication Department, Bengal Engineering and Science University, India.*

mized value of µ is computed, to reduce peak to peak power ratio (PAPR) and to improve the efficiency of the multicarrier modulation technique like OFDM. Also companding filter is proposed in [11], [12], and [13]. In [14] Roy et.al presents a generalized synthesis methodology of including compander, using OTA. The above discussion depicts that though the attention has been given to design compander circuit using digital components, to avoid expensive and bulky analog discrete communication circuit, components and higher power supply requirement, or to optimize the value of µ to reduce PAPR. In this scenario, this work presents design methodology of communication (analog and digital) circuits (Delta Modulator and Compander) with an array of OTAs and active element like semiconductor diode. The simulation results of the circuit performance have also been reported. Section 2 and 3 reports the design methodology of OTA based radio frequency non linear communication circuits. Finally, the simulation results are highlighted in section 4 to verify the performance of the designed circuits.

## 2 DESIGN OF DELTA MODULATOR

This section deals with design methodology of OTA based linear and Adaptive Delta Modulator circuit used in communication system.

### 2.1 Delta Modulator

Delta modulation (DM) may be viewed as two-level (1-bit) quantizer. A block diagram of DM encoder is shown in Fig. 1. From this figure, we obtain

$$m_q[k] = m_q[k-1] + d_q[k] \tag{1}$$

Hence,

$$m_q[k-1] = m_q[k-2] + d_q[k-1] \tag{2}$$

Substitution of Eqn. 1 in Eqn. 2 yields

$$m_q[k-1] = m_q[k-2] + d_q[k-1] \tag{3}$$

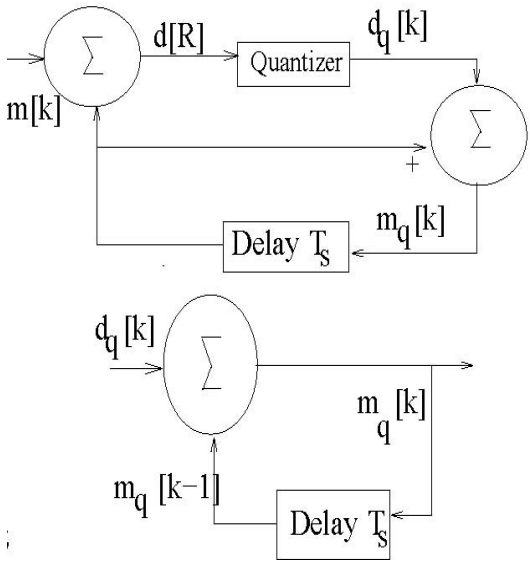

Fig. 1. Block diagram of Delta Modulator.

Proceeding iteratively and assuming zero initial condition, (i.e., $m_q[-1] = 0$ and $d_q[0] = 0$) we arrive at

$$m_q[k] = \sum_{m=0}^{k} d_q[m] \tag{4}$$

Equation 4 represents an accumulator (adder). If the output $d_q[k]$ is represented by impulses, then the accumulator may be realized by an integrator. OTA network of linear delta modulator is shown in Fig. 2. In that figure (Fig. 4) the portion of the circuit encircled by dashed lines represent OTA realization for the linear delta modulator. Along with that portion, circuit outside the dashed line represents continuously variable slope delta modulator (discussed in the next section). Subsequent discussions analyze linear delta modulator. Along with that portion, circuit outside the dashed line represents continuously variable slope delta modulator (discussed in the next section). OTA1 and OTA2 serve the purpose of a deference amplifier leading to a two level quantizer. The output of the quantizer assumes the supply voltage level +V and -V depending on whether the input overrides the pre-dicted signal or not. The output of the quantizer is sampled at a rate $f_s = 2kf_m$ (by the switch $S_1$ in Fig. 2), where $f_m$ is the frequency of the input sine wave and k is the oversampling factor. The sampled quantizer output is fed next to the input of an integrator implemented by the OTA3 and capacitor C1. The input to the integrator is given by the following equation.

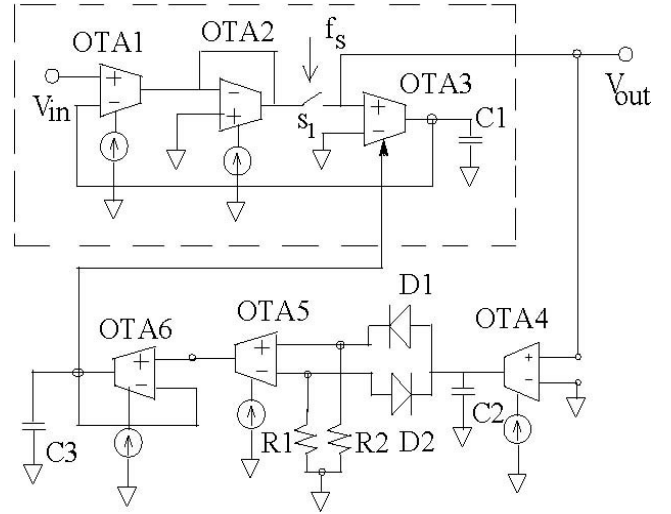

Fig. 2. OTA realization of Adaptive Delta Modulator.

$$s_i(t) = V \sum a_n p(t - nT_s) \tag{5}$$

where V is the maximum output level of the quantizer and $a_n$ is either +1 or -1, and $T_s = 1/f_s$. The pulse shape p(t) is given by

$$p(t) = u(t) - u(t - \frac{T_s}{2}) \tag{6}$$

where u(t) is Heaviside step function. Assuming ideal OTA, the output of the integrator is given by

$$s_0(t) = \frac{g_{m3}}{C_1} \sum a_n p(t - nT_s) - \rho(t - nT_s - \frac{T_s}{2}) \quad (7)$$

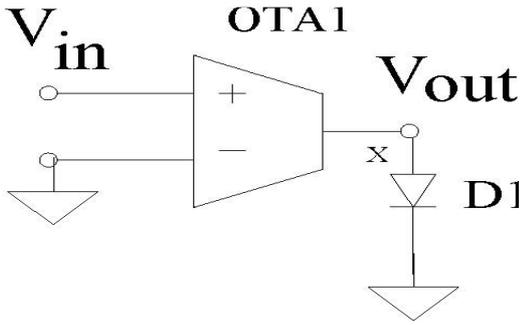

Fig. 3. OTA realization of Compressor.

Where $\rho(t)$ is a ramp function and $T_s = 1/f_s$.
The step size $\delta$ of the staircase waveform is given by

$$\delta = \frac{g_{m3} T_s}{2C_1} \quad (8)$$

## 2.3 Continuously Variables slope Delta Modulator

A linear delta modulator requires a large over sampling factor 'k' (from 8 to 16) for proper operation. By operating the modulator in an adaptive mode which in it in the granular region the over-sampling factor can be reduced down to 4 to 8 thus decreasing the output bit rate. A continuously variable slope delta modulator adapts the step size in a continuous fashion both in the granular and overload regions. Fig. 2 (including linear delta modulator encircled by dashed lines) represents continuously variable slope delta modulator. The front end of the circuit (encircled by dashed line) is an integrator followed by a rectifier and a low pass filter (LPF). When the delta modulator is overloaded the input to the integrator will be a sequence of all positive or all negative pulses. The magnitude of the output of the integrator in this case will be higher. The integrator output is rectified and filtered by LPF. The output of the low pass filter increases the transconductance $g_m3$ of the integrator of the linear delta modulator block. As a result step size (δ) increases because step size (δ) directly varies with $g_m3$ (equation 8). When the modulator hunts around a flat portion of the input signal the output of the switch will be alternate in polarity (i.e. its magnitude changes between positive and negative values alternatively). As a result, transconductance of OTA3 ($g_m3$) will assume a low value. Consequently, from equation (8) it can be concluded that step size (δ) decreases. Thus Fig. 4 realizes continuously variable slope delta modulator increases the step size in the overload region and decreases

## 3 COMPANDER

For transmission of speech signal using waveform coding techniques the same quantizer has to accommodate input signals with widely varying power level. The range of voltages covered by voice signals, from loud talk to weak talk, is in the order of 1000 to 1. For acceptable voice transmission signal-to-quantization noise ratio should remain essentially constant for wide range of input power levels. If the step size of the quantizer increases non-uniformly then the dynamic range of the quantizer is improved. By using a device called compressor the desired form of the non-uniform quantization can be achieved. Characteristic of a compressor can be mathematically represented by logarithmic function [15]. Figure 3 shows OTA realization of compressor. Analysis of OTA-Compressor circuit is as follows. At the output node of OTA2 of that circuit, characteristic of a compressor can be expressed by logarithmic

$$v_0(t) = \frac{\ln(1 + \mu \cdot v_i(t))}{\ln(1 + \mu)} \quad (9)$$

where $v_i(t)$ and $v_o(t)$ are the magnitude of the input and output signal respectively and μ is a parameter that is selected to give the desired compress characteristic. If it is assumed $\ln(1+\mu) = D$, then the equation (1) can be written as, $e^{D \cdot v_0(t)} - 1 = \mu \cdot v_i(t) \quad (10)$

A semiconductor diode is characterized by the equation

$$I = I_s(e^{\frac{V \cdot q}{nKT}} - 1) \quad (11)$$

where $I_s$ = reverse saturation current, n=emission coefficient, K is the Boltzman constant (K=1.381x10⁻²³), T is the absolute temperature in degree Kelvin, and q is the charge of an electron (q =1.6x10⁻¹⁹). The symbol 'n' is unity for germanium and approximately 2 for silicon [16]. Therefore exponential behavior of equation (10) conceived vividly as real by semiconductor p-n junction diode equation [(3)] and an OTA. Fig. 3 displays the OTA based compressor circuit. Applying K.C.L at the node 'x' in the Fig. 3 we get,

$$V_{in} \cdot g_m = (e^{\frac{V_{out} g}{nKT}} - 1) I_s \quad (12)$$

$$e^{\frac{qV_{out}}{nKT}} = 1 + \ln[1 + \frac{V_{in} \cdot g_m}{I_s}] \quad (13)$$

Taking natural logarithm on both side of the equation (11) produces

$$\frac{q \cdot V_{out}}{nKT} = \ln[1 + \frac{V_{in} \cdot g_m}{I_s}] \quad (14)$$

Thus the output of the compressor is stated as,

$$V_{out} = \frac{nKT}{q}[\ln(\frac{V_{in} \cdot g_m}{I_s} + 1)] \quad (15)$$

At room temperature 300 K the value of $\frac{KT}{q}$ =26mV and assuming n=2, output of the compressor can be written as,

$$\frac{V_{out}}{x} = \frac{26 \times 10^{-3}}{x} \ln[1 + \frac{V_{in} \cdot g_m}{I_s}] = \frac{\ln[1 + \frac{V_{in} g_m}{I_s}]}{x \times 19.23} \quad (16)$$

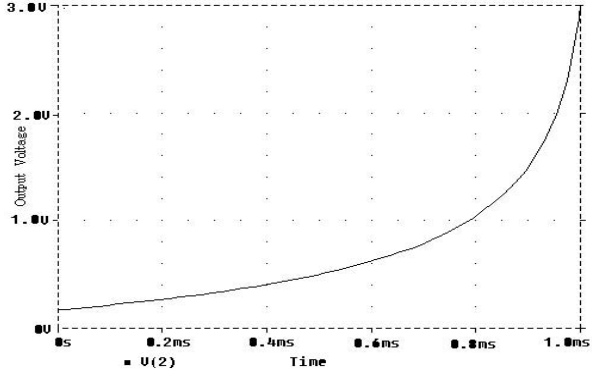

Fig. 6. Output characteristic of simulated Expander.

By comparing eqn (9) and (15) it can be stated that $\frac{g_m}{I_s}$ is equivalent to $\mu$ and $x \times 19.23 = \ln[1+\mu]$. Thus the value of $\mu$ can tuned with the transconductance of OTA

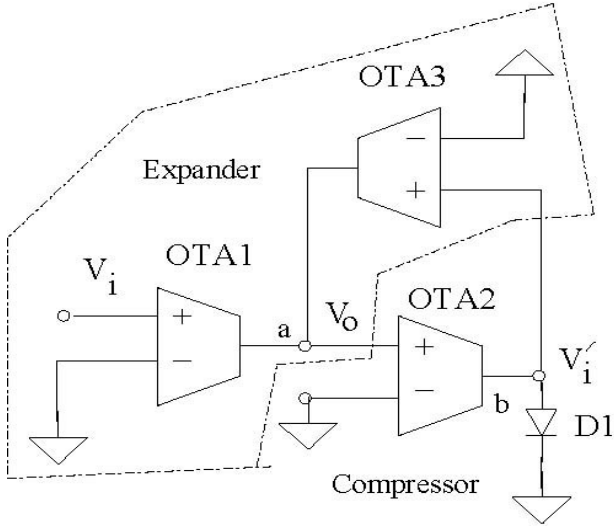

Fig. 4. OTA realization of Compander.

i.e. $g_m$.

Inverse process of compressor is known as expander. Thus expander output can be expressed as

$$v_0(t) = e^{v_i(t)} \quad (17)$$

$v_i(t)$ is the compressed signal and input of the expander and $v_0(t)$ is the output of the expander. Fig. 4 depicts the OTA based expander circuit. At node 'b' in Fig. 6 applying KCL we get,

$$g_{m2}V_0 = (e^{\frac{V_i'q}{KT}} - 1)I_s \quad (18)$$

Hence

$$V_i' = \frac{\ln[1+\frac{V_0 g_{m2}}{I_s}]}{38.46} \quad (19)$$

Equation (18) is valid at room temperature 300K (ref. equation (15). Again applying K.C.L at node 'a', we get

$$V_i . g_{m1} = V_i' g_{m3} \quad (20)$$

Substituting values of $V_i'$ from equation (18) in equation (17), and putting $\mu = \frac{I_s}{g_{m3}}$ we get

$$V_0 = \frac{1}{\mu}[e^{\frac{g_{m1}}{g_{m3}}V_i . 38.46} - 1] \quad (21)$$

## 4 SIMULATION RESULTS AND DISCUSSION

To confirm the practical validity of the proposed circuits, shown in Fig 2, 3 and 4 are simulated using CADENCE SPICTRE cad tool. A CMOS OTA structure with dimension of the channel length (L) and width (W) of the NMOS and PMOS are 0.3µm and 0.4µm respectively. Parasitic output capacitance of the OTA is 2.1 fF. A simple CMOS linear OTA with deferential input was proposed by Szczepaski [13]. Such an OTA with $g_m$ varying 70 to 120 µA/V is employed in our design application of Delta Modulator. Figure 7 shows the input and output signal of the delta modultor. Frequency of the input signal is 10 MHz and step size is 250 mv. Step size can be tuned by varying OTA3 and C1 in Fig. 4. The output of adaptive delta modulator is depicted in Fig. 8. Like delta modulator the input signal frequency is 10 MHZ and the sampling frequency is 90 MHZ. It is clear from Fig. 8 that step size varies from 0.4 volt to 0.9 volt. Fig. 5 depicts the simulated output characteristic of compressor. As the practical value of µ of equation (1) is approximately 255 the range of 10µA to 50µA [10], thus the transconductance of the OTA is tuned from 1mA/V to 50mA/V.

TABLE 1: VALUES OF

$g_m$, $I_s$ and µ

| $g_m$ | $I_s$ | µ |
|---|---|---|
| 10 mA/V | 28.5µA | 350 |

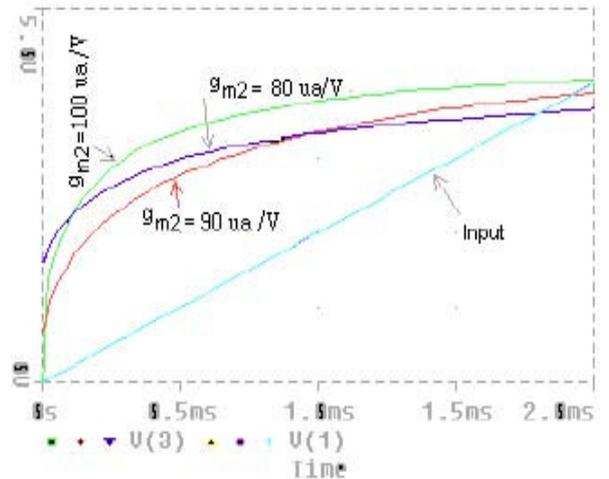

Fig. 5. Output characteristic of simulated Compressor.

| 9 mA/V | 39.5 µA | 230 |
|---|---|---|
| 8 MA/V | 39.5 µA | 204 |

A ramp signal, which varies from 0V to 2.5V during the time interval 0 to 1ms, is applied at the input of the compressor. Output characteristic curves depicts for three different values of $g_m$, $I_S$, and µ as shown in the Table 1. Fig. 5 shows the simulated output of the expander. It can be shown that the simulated result validates the theory of the compressor discussed in Section 3.

## 5 CONCLUSION

The potential of the use of current mode device, OTA, for RF circuit has been discussed. From considerations of cell based design of non linear circuit, OTA is preferred for its simplicity. Though three types of communication circuits are presented in this paper, using OTA as basic building

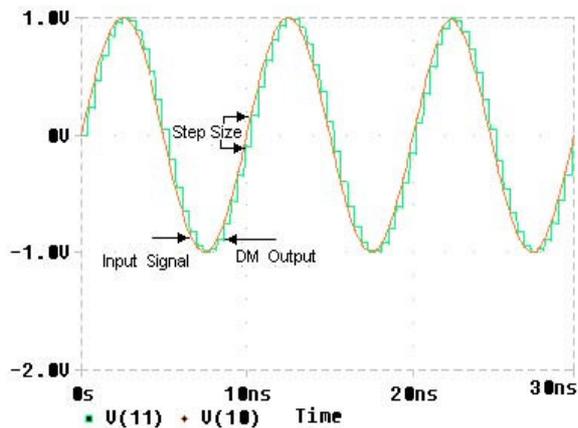

Fig. 7. Output characteristic of Delta Modulator (DM).

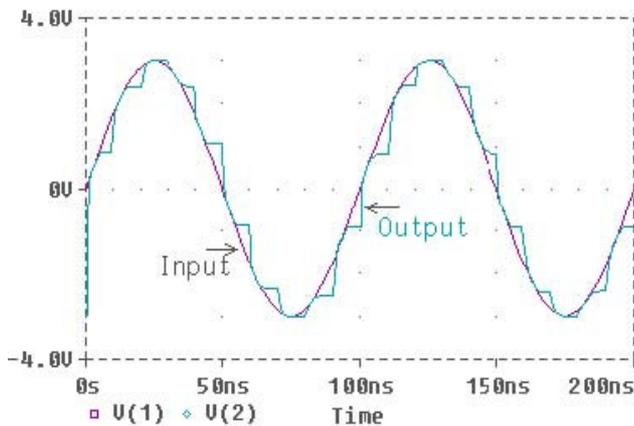

Fig. 8. Output characteristic of Adaptic Delta Modulator (ADM).

block and with the cell library concept other types of nonlinear circuits can easily be designed.
.